\documentclass{article}
\usepackage{romp}
\usepackage{amsmath}
\usepackage{amssymb}
\usepackage{graphicx}
\usepackage[numbers,sort&compress]{natbib}

\usepackage{amscd, amsfonts, amssymb}
\usepackage{epsfig}
\usepackage[mathscr]{eucal}
\usepackage{verbatim}
\usepackage{latexsym}
\usepackage{lscape}
\usepackage{enumerate}

\newcommand\ra{\rightarrow}

\newcommand{\ovl}[1]{{\overline{#1}}}

\newcommand{\nfs}{{}}

\def\e{{\mathrm{e}}}

\title{Dark matter, {E}lko fields and {W}einberg's quantum field theory formalism}
\author{Adam Gillard
\\Department of Physics and Astronomy,
University of Canterbury,
Private Bag 4800, \hspace*{\fill}\\
Christchurch 8140,
New Zealand \hspace*{\fill}\\Email: adam.gillard@pg.canterbury.ac.nz  \\[2ex]
         Benjamin Martin\thanks{\ Corresponding author}
                      \\Department of Mathematics,
University of Auckland,
Private Bag 92019, \hspace*{\fill}\\
Auckland 1142,
New Zealand  \hspace*{\fill}\\Email: Ben.Martin@auckland.ac.nz }

\begin{document}

\maketitle
\begin{abstract}
     The Elko quantum field was introduced by Ahluwalia and Grumiller, who proposed it as a candidate for dark matter.
     We study the Elko field in Weinberg's formalism for quantum field theory.  We prove that if one takes the symmetry group to be the full Poincar\'e group then the Elko field is not a quantum field in the sense of Weinberg.  This confirms results of Ahluwalia, Lee and Schritt, who showed using a different approach that the Elko field does not transform covariantly under rotations and hence has a preferred axis.
\end{abstract}


\section{Introduction}
\label{sec:intro}

Experimental evidence strongly suggests that much of the matter in the universe is hidden from us; observed phenomena such as the temperature distribution of hot gas in galaxies and clusters of galaxies, orbital velocities of galaxies in clusters, and the rotational speeds of galaxies, all give evidence of this missing mass.  The existence of so-called dark matter was postulated by Fritz Zwicky to explain the motion of the Coma cluster of galaxies. Using the virial theorem, he found that the galaxies should have about 400 times the amount of mass than was visually observed \cite{Zwi33}.  The missing matter is referred to as {\em dark matter}.  Dark matter interacts only very weakly with Standard Model matter and electromagnetic radiation.  Many attempts have been made to give a theoretical construction of dark matter and to explain the mechanisms that suppress its interactions with Standard Model particles \cite{bacon2000detection}\cite{hinshaw2009five}\cite{peccei1977constraints}\cite{weinberg1978new}.

Ahluwalia and Grumiller in 2005 constructed a new quantum field, which they named the {\em Elko field} \cite{AG05}.  It is a spin-half fermionic quantum field of mass dimension one.  The original version of the field is non-local; a modified local version was constructed by Ahluwalia, Lee and Schritt \cite{ahluwalia2010elko}.  Ahluwalia and Grumiller proposed the Elko field as a candidate for dark matter.  The Elko field has been the subject of much attention
\cite{wei2010couple}\cite{shankaranarayanan2009if}\cite{fabbri2010causal}\cite{fabbri2010causal2}\cite{FBZEOPWFE}\cite{dRR11}.
For instance, da Rocha and Rodrigues showed in 2006 that the Elko spinor fields belong to the class of flagpole spinor fields in the Lounesto classification \cite{dRR06}.  Indeed, da Rocha and Hoff da Silva defined a mapping from Dirac spinor fields to Elko spinor fields with the aim of extending the Standard Model to spinor fields of mass dimension one.  In \cite{dRR09} they define a map between the Dirac and Elko Lagrangians, after defining in \cite{dRR07} the transformations mapping Dirac to Elko spinor fields.  In \cite{dRR08} this transformation of fields is shown to be analogous to the instanton map arising from a Hopf fibration.  Da Rocha and Hoff da Silva used Elko fields to derive the Einstein-Hilbert-, Einstein-Palatini- and Holst actions from the quadratic spinor Lagrangian \cite{dRR09a}.

Ahluwalia, Lee and Schritt noted in late 2009 that the Elko field does not transform covariantly under rotations: they showed that the Elko spin sums contain a preferred direction \cite[Sec.~2.5]{ahluwalia2010elko}, \cite[Sec.~IID~and~Sec.~IV]{ALS09b}, which implies that rotational symmetry is violated.  See also \cite[Sec.~3.1]{BBMS2010}.  This result was foreshadowed in \cite[p10]{Ahl03}.  Ahluwalia and Horvath \cite{AH} constructed a variant of the Elko field having as its symmetry group the group SIM(2), which arises in the theory of Very Special Relativity \cite{CG}.  Experimental measurements of the cosmic microwave background support the existence of a preferred axis for dark matter---the so-called ``axis of evil'' \cite{LM05}\cite{LM07}.  A striking feature of the Elko field is that the preferred axis is built into the theory, rather than having to be imposed externally.

In this paper we show that if one assumes the symmetry group to be the Poincar\'e group then the Elko field is not a quantum field in the sense of Weinberg \cite[Ch.~5]{Wei95}.  Ahluwalia et al.\ take the usual spinorial approach to field theory: they specify a field by specifying the associated spinors.  (Indeed, the original motivation for the Elko field was to construct a field for which the spinors are eigenspinors of the charge conjugation operator.)  Our work complements theirs: we take a quantum-field-theoretic approach, working within the formalism for quantum field theory introduced by Weinberg \cite{Wei95}.  Giving the data for a Weinberg quantum field involves explicitly specifying a Hilbert state space $H$, a unitary representation $U(\Lambda,a)$ of the Poincar\'e group ${\mathcal P}$ on $H$ and a finite-dimensional representation $D(\Lambda)$ of the Lorentz group ${\mathcal L}$; the formulas for the spinors are not taken as given but are to be derived from the transformation properties of the field \cite[Sec.~5.1]{Wei95}.  We give a rigorous proof within Weinberg's formalism that the Elko field does not transform covariantly under the whole of the Poincar\'e group, thereby confirming the conclusions of \cite[Sec.~2.5]{ahluwalia2010elko} and reinforcing the necessity for a preferred frame analysis.

We start by considering the usual $H$, $U(\Lambda,a)$ and $D(\Lambda)$.  It is immediate that the Elko field is not a Weinberg quantum field for this data because of Weinberg's result \cite[Sec.~5.5]{Wei95} that the Dirac field is the only such field.  We then prove, however, that even if one changes the usual representations of the Poincar\'e and Lorentz group by applying similarity transforms, one still cannot obtain a Weinberg quantum field of Elko type; this is much less obvious.  This shows that the Elko field is not merely the Dirac field in disguised form, but is genuinely new.  We obtain our results by studying the form of the Elko rest spinors.  These results apply not just to the particular Elko rest spinors chosen in \cite{AG05} and \cite{ahluwalia2010elko}, but to a more general class of rest spinors, those of the form given in eq.~(\ref{eqn:genelko}).

Our paper has two main purposes.  The first is to contribute to the theory of Elko fields.  The second is to illustrate the value of Weinberg's formalism as a complement to the usual spinorial approach to quantum field theory.  Weinberg's formalism gives a systematic way of looking at quantum field theory relying on a minimum of physical assumptions.  When the transformation properties of a field are discussed, there is sometimes scope for ambiguity because there are actually two group representations involved: the unitary representation of the Poincar\'e group on the state space, and the finite-dimensional representation of the Lorentz group on the components of the field.  Weinberg's formalism makes this structure very explicit; his criterion for a collection of spinors to form a quantum field says roughly that the two representations should behave compatibly (eq.~(\ref{eqn:fieldeqn})).  This is a purely mathematical requirement, so the possible fields can be determined (at least in principle) without invoking any extra physical assumptions: the physics falls out as an inevitable consequence of the representation theory.  One may also, for example, deduce the existence of anti-particles from eq.~(\ref{eqn:fieldeqn}) together with the requirement of locality \cite[p199]{Wei95}.

Like all theories, the formalism of Weinberg presented in \cite[Ch.~5]{Wei95} has its limitations.  It is strictly a special relativistic theory: it does not incorporate gravitational effects or allow for curved space-times\footnote{Note that there has been much recent research on Elko theory in curved space-times.  B\"ohmer showed that Elko spinors couple naturally to torsion \cite{boehmer2007einstein1}\cite{Boe07}.  Some theories involving nonlinear Elko interactions are described in \cite{F1}\cite{F2}\cite{F3}.  For work in a supersymmetric setting, see \cite{WD}.}.  The possible interaction terms $V$ are restricted \cite[p110]{Wei95} and topologically twisted fields such as magnetic monopoles are ruled out \cite[p119]{Wei95}\footnote{The framework of Elko spinor fields can deal with topologically twisted fields and seems suitable eventually to probe non-trivial space-time topologies \cite{dRBH1}\cite{dRBH2}.}.  Nonetheless we believe that if these limits are respected then Weinberg's formalism is a valuable tool both for constructing quantum fields and for determining obstructions to their existence.  We hope this paper will help stimulate further work on this and other foundational approaches to quantum field theory.

Some of our results were announced at the Dark2009 conference in January 2009 and an abbreviated form of this paper appeared in the conference proceedings \cite{Conferenceproceedings}.

The paper is laid out as follows.  In Section~\ref{sec:reps} we briefly recall the Poincar\'e and Lorentz groups and their representations.  Section~\ref{sec:elkos} contains a review of the most important aspects of Elko fields, based on \cite{AG05} and \cite{ahluwalia2010elko}.  In Section~\ref{sec:weinberg} we review Weinberg's theory and in Section~\ref{sec:diracelko} we study the Dirac and Elko fields in Weinberg's framework and establish our main result.  We finish with a brief discussion of non-standard Wigner classes in Section~\ref{sec:nonstd} \nfs  Our notation is standard, and for the most part follows that of Weinberg \cite{Wei95}.


\section{Representations of the Poincar\'e and Lorentz groups}
\label{sec:reps}

It is a basic physical principle that symmetries of a physical system give rise to unitary or anti-unitary maps on the state space $H$.  This yields a representation\footnote{By a representation of a group $G$ on a complex vector space $V$ we mean a projective corepresentation: that is, a function $f$ from $G$ to ${\rm GL}^\#(V)$ satisfying the homomorphism property $f(g_1g_2)= f(g_1)f(g_2)$ up to phase.  Here ${\rm GL}^\#(V)$ is the group (under composition) of invertible functions $T\colon V\ra V$ such that $T$ is either linear or anti-linear.}
of the symmetry group on $H$.  If this representation is reducible then $H$ splits up into a direct sum of subrepresentations which don't interact with each other, so it is usually enough to consider irreducible representations.  We recall some facts about representations of the Poincar\'e and Lorentz groups.

We denote by ${\mathcal L}^0$ the connected component of the Lorentz group, which consists of the proper orthochronous Lorentz transformations, and we call ${\mathcal L}^0$ the {\em strict Lorentz group}.  Then ${\mathcal L}^0$ is generated by rotations and boosts.  By the {\em Lorentz group} ${\mathcal L}$ we mean the extended Lorentz group, including the discrete symmetries of space inversion ${\sf P}$ and time reversal ${\sf T}$ and their product ${\sf P}{\sf T}$.  Likewise we use the term {\em Poincar\'e group} to refer to the extended Poincar\'e group ${\mathcal P}$ generated by ${\mathcal L}$ together with space-time translations, and the term {\em strict Poincar\'e group} to refer to the connected component ${\mathcal P}^0$ of ${\mathcal P}$, which is generated by ${\mathcal L}^0$ together with space-time translations.  Usually we use the symbol $\Lambda$ to denote an element of ${\mathcal L}$, and we represent elements of ${\mathcal P}$ by pairs $(\Lambda,a)$, where $\Lambda\in {\mathcal L}$ and $a= a^\mu$ is a space-time translation.

The finite-dimensional irreducible representations of ${\mathcal L}^0$ are classified up to isomorphism by integer- or half-integer pairs: for instance, $(\frac{1}{2},\frac{1}{2})$ corresponds to the vector representation and $(\frac{1}{2},0)\oplus(0,\frac{1}{2})$ corresponds to the chiral representation, which is reducible.  In Section~\ref{sec:elkos} we briefly recall an explicit construction of the chiral representation, which we need in our discussion of the Dirac and Elko fields.

Now we consider the irreducible unitary representations of ${\mathcal P}^0$.  These were first constructed by Wigner \cite{Wig39}; a thorough account is given in \cite[Ch.~2]{Wei95}.  Fix $m>0$ and a non-negative integer or half-integer $s$.  Denote the infinitesimal generators of space-time displacement by $P^\mu$ and the infinitesimal generators of the Lorentz group by $J^{\mu \nu}$.  The $P^\mu$ and the $J^{\mu \nu}$ span the Lie algebra of ${\mathcal P}^0$.  We obtain a Hilbert space $H_1$ and an irreducible unitary representation $U(\Lambda,a)$ of ${\mathcal P}^0$ as follows.  There is a basis of vectors denoted $\left|p,\sigma\right\rangle$, where $p= p^\mu$ satisfies $p_\mu p^\mu= -m^2$, $p^0>0$ and $\sigma\in \{-s,-s+1,\ldots, s-1,s\}$.  The inner product on $H_1$ is given by
\begin{eqnarray}
\left\langle p',\sigma'|p,\sigma\right\rangle=\delta^3(\mathbf{p}'-\mathbf{p})\delta_{\sigma'\sigma}.
\end{eqnarray}

Let $L(p)$ be the Lorentz boost that takes the rest frame to the frame with 4-momentum $p$.  Given $\Lambda\in {\mathcal L}^0$, define $W(\Lambda,p)$ by

\begin{equation}
 W(\Lambda,p)= L^{-1}(\Lambda p)\Lambda L(p).
\end{equation}

\noindent Let $k$ be the 4-vector $(p^0,0,0,0)$.  Note that each $W(\Lambda,p)$ fixes the vector $(p^0,0,0,0)$, so $W(\Lambda,p)$ belongs to the subgroup ${\rm SO}(3)$ of ${\mathcal L}^0$---the so-called {\em little group}.

The operators $U(\Lambda,a)$ of the representation are given by

\begin{eqnarray}
U(\Lambda,a)\left|p,\sigma\right\rangle=\sqrt{\frac{(\Lambda p)^0}{p^0}}\e^{-i(\Lambda p)_{\mu}a^{\mu}}\sum_{\ovl{\sigma}}R_{\ovl{\sigma}\sigma}(W(\Lambda,p))\left|\Lambda p,\ovl{\sigma}\right\rangle.
\end{eqnarray}

\noindent Here $R(\Lambda)$ is an irreducible spin-$s$ representation of ${\rm SO}(3)$ (note that there is exactly one such representation up to similarity transform).  In particular, the vectors $\left|k,\sigma\right\rangle$ span an ${\rm SO}(3)$-stable subspace of $H_1$, which transforms according to the representation $R(\Lambda)$:

\begin{equation}
\label{eqn:spin}
 U(\Lambda)\left|k,\sigma\right\rangle= \sum_{\ovl{\sigma}}R_{\ovl{\sigma}\sigma}(\Lambda)\left|k,\ovl{\sigma}\right\rangle,
\end{equation}

\noindent for $\Lambda\in {\rm SO}(3)$.  It follows that

\begin{equation}
\label{eqn:spinangmom}
 J_z\left|k,\sigma\right\rangle=\sigma\left|k,\sigma\right\rangle,\quad
\boldsymbol{J}^2\left|k,\sigma\right\rangle=s(s+1)\left|k,\sigma\right\rangle.
\end{equation}

\noindent The $\left|p,\sigma\right\rangle$ are eigenvectors of $P^{\mu}$: we have

\begin{equation}
\label{eqn:Pevector}
 P^{\mu}\left|p,\sigma\right\rangle=p^{\mu}\left|p,\sigma\right\rangle.
\end{equation}

The usual physical interpretation holds: $H_1$ is the state space of a single particle of mass $m$ and spin $s$, and the ket $\left|p,\sigma\right\rangle$ represents a particle with 4-momentum $p$ and rest-frame spin $\sigma$ in a direction $\widehat{\mathbf n}$ which is conventionally chosen to be the $z$-direction.  We consider only the case of positive mass.

In Weinberg's derivation, one starts with basis vectors $\left|p,\sigma\right\rangle$ satisfying eq.~(\ref{eqn:Pevector}), where $\sigma$ is assumed only to be a discrete index labelling all the remaining degrees of freedom \cite[p63]{Wei95}.  The derivation shows that eq.~(\ref{eqn:spin}) must hold, for some irreducible unitary representation $R(\Lambda)$ of ${\rm SO}(3)$.  Hence the index $\sigma$ is forced to take on the values $-s,-s+1,\cdots s-1,s$ for some integer or half-integer $s$.

To do quantum field theory one needs to allow for states with several particles.  Consider a particle of species $n$.  One constructs the multi-particle state space $H_{\textrm{tot},n}$ from the one-particle state space $H_{1,n}=H_1$ by taking a direct sum of symmetric or anti-symmetric tensor powers of $H_{1,n}$, depending on whether the particles concerned are bosons or fermions. One introduces creation operators $a^{\dagger}(\mathbf{p},\sigma)$ and annihilation operators $a(\mathbf{p},\sigma)$ in the usual way.
Finally, one forms the total state space $H=H_{\textrm{tot}}$ as the direct sum over all the particle species of the state spaces $H_{\textrm{tot},n}$. Then $H$ inherits a unitary representation of the Poincar\'e group from the representation on each $H_{1,n}$: we abuse notation and write $U(\Lambda,a)$ for this representation as well.  We refer the reader to \cite[Ch.~2 and Ch.~4]{Wei95} for details.

As a final remark in this section, we note that one should be careful with the term ``spin'' in a relativistic setting. The appropriate relativistic notion is given by using the Casimir operator $W_{\mu}W^{\mu}$, where $W_\mu= -\frac{1}{2}\epsilon_{\mu\nu\rho\sigma}J^{\nu\rho}P^{\sigma}$ is the Pauli-Lubanski operator.  We have

$$ \quad W_{\mu}W^{\mu}\left|p,\sigma\right\rangle=-m^2s(s+1)\left|p,\sigma\right\rangle. $$

\noindent In the rest frame, $W_\mu W^\mu$ is a multiple of $\boldsymbol{J}^2$: we have

\begin{equation}
 W_\mu W^\mu \left|k,\sigma\right\rangle= -m^2J_\mu J^\mu \left|k,\sigma\right\rangle= -m^2\boldsymbol{J}^2 \left|k,\sigma\right\rangle= -m^2s(s+1)\left|k,\sigma\right\rangle
\end{equation}
for any $\sigma$.

\section{Review of the Elko field}
\label{sec:elkos}

The underlying finite-dimensional representation of ${\mathcal L}^0$ for both the Elko field and the Dirac field is the chiral (or Weyl bispinor) representation $D^{\rm ch}(\Lambda)$, which is of type $(\frac{1}{2},0)\oplus(0,\frac{1}{2})$.  We recall the construction of this arising from the chiral representation of the gamma matrices.

Define the gamma matrices by

\begin{eqnarray}
\gamma^0=\left(\begin{array}{cccc}
0&\mathbb{I}\\
\mathbb{I}&0
\end{array}\right),\quad
\gamma^i=\left(\begin{array}{cccc}
0&-\sigma_i\\
\sigma_i&0
\end{array}\right),
\end{eqnarray}

\noindent where $\mathbb{I}$ denotes the $2\times 2$ identity matrix and the $\sigma_i$ are the Pauli matrices given by

\begin{eqnarray}
\sigma_1=\left(\begin{array}{cccc}
0&1\\
1&0
\end{array}\right),\quad \sigma_2=\left(\begin{array}{cccc}
0&-i\\
i&0
\end{array}\right)\quad\sigma_3=\left(\begin{array}{cccc}
1&0\\
0&-1
\end{array}\right).
\end{eqnarray}

\noindent We define

\begin{equation}
 J^{\mu \nu}= -i[\gamma^\mu,\gamma^\nu];
\end{equation}

\noindent this yields a representation of ${\rm Lie}\,{\mathcal L}^0$.  Exponentiating yields a representation of ${\mathcal L}^0$.  For example, the operator corresponding to a Lorentz boost $L(p)$ is given by $\kappa= \kappa(\mathbf{p})= \kappa^+\oplus\kappa^-$ where

\begin{eqnarray}
\kappa^+&=&\sqrt{\frac{E+m}{2m}}\left(\mathbb{I}+\frac{\boldsymbol{\sigma}\cdot\mathbf{p}}{E+m}\right),\\
\kappa^-&=&\sqrt{\frac{E+m}{2m}}\left(\mathbb{I}-\frac{\boldsymbol{\sigma}\cdot\mathbf{p}}{E+m}\right).
\end{eqnarray}

\noindent We define $P$ to be the parity operator

\begin{equation}
 P= \left(\begin{array}{cccc}
 0&\mathbb{I}\\
 \mathbb{I}&0
 \end{array}\right)
\end{equation}

\noindent and $C$ to be the charge conjugation operator

\begin{eqnarray}
C=\left(\begin{array}{cccc}
\mathbb{O}&i\Theta\\
-i\Theta&\mathbb{O}
\end{array}\right)K,
\end{eqnarray}

\noindent where $\mathbb{O}$ is the $2\times 2$ zero matrix, $\Theta$ is the Wigner spin-half time reversal operator given by

\begin{eqnarray}
\Theta=\left(\begin{array}{cccc}
0&-1\\
1&0
\end{array}\right)
\end{eqnarray}

\noindent and $K$ is the antilinear operator that acts by complex conjugation to the right.

In \cite{AG05}, Ahluwalia and Grumiller proposed a non-local, mass dimension one spin-half quantum field $\eta(x)$. We shall consider the more recent local modified field $\Lambda(x)$ given in \cite{ahluwalia2010elko}.
First we recall the definition of the Dirac field \cite[eq.~(5.5.34)]{Wei95}, which is given by\footnote{The Dirac field is often written with a factor of $\dfrac{1}{\sqrt{2E(\mathbf{p})}}$ in the integrand.  We have absorbed this factor into the definition of the spinors $u$ and $v$; cf.\ eqs.~(\ref{eqn:uboost}) and (\ref{eqn:vboost}).}

\begin{equation}
\label{eqn:diracfld}
\psi(x)=\int \frac{d^3\mathbf{p}}{(2\pi)^{\frac{3}{2}}}
\sum_{\sigma}\left[\e^{ip^\mu x_\mu}u(\mathbf{p},\sigma)a(\mathbf{p},\sigma)+\e^{-ip^\mu x_\mu}v(\mathbf{p},\sigma)b^{\dagger}(\mathbf{p},\sigma)\right],
\end{equation}

\noindent where the Dirac rest spinors $u(\mathbf{0},\sigma)$ and $v(\mathbf{0},\sigma)$ are defined by

\begin{eqnarray}
\label{eqn:diracrestA}
u(\mathbf{0},{\textstyle \frac{1}{2}})=\frac{1}{\sqrt{2}}\left(\begin{array}{cccc}
1\\
0\\
1\\
0
\end{array}\right),\quad u(\mathbf{0},-{\textstyle \frac{1}{2}})=\frac{1}{\sqrt{2}}\left(\begin{array}{cccc}
0\\
1\\
0\\
1
\end{array}\right),\\
\label{eqn:diracrestB}
v(\mathbf{0},{\textstyle \frac{1}{2}})=\frac{1}{\sqrt{2}}\left(\begin{array}{cccc}
0\\
1\\
0\\
-1
\end{array}\right),\quad v(\mathbf{0},-{\textstyle \frac{1}{2}})=\frac{1}{\sqrt{2}}\left(\begin{array}{cccc}
-1\\
0\\
1\\
0
\end{array}\right)
\end{eqnarray}

\noindent and

\begin{eqnarray}\label{eqn:uboost}
 u(\mathbf{p},\sigma) & = & \sqrt{\frac{m}{E(\mathbf{p})}}\; \kappa u(\mathbf{0},\sigma),
\end{eqnarray}
\begin{eqnarray}\label{eqn:vboost}
 v(\mathbf{p},\sigma) & = & \sqrt{\frac{m}{E(\mathbf{p})}}\; \kappa v(\mathbf{0},\sigma),
\end{eqnarray}

\noindent with $\kappa$ as above \cite[eqs.~(5.5.6) and (5.5.7)]{Wei95}.  The Dirac rest spinors are eigenspinors with eigenvalue $\pm 1$ of the parity operator $P$.

The local Elko quantum field \cite[eq.~(38)]{ahluwalia2010elko} is a four-component spinor field given by\footnote{We have changed the signs in the exponentials of eq.~(\ref{eqn:localelko}) to fit in with the conventions of \cite{Wei95}: this amounts simply to adopting a different convention in the definition of how the translation operators act on state vectors.}
\begin{equation}
\label{eqn:localelko}
\Lambda(x)=\int \frac{d^3\mathbf{p}}{(2\pi)^3}\frac{1}{\sqrt{2mE(\mathbf{p})}} \sum_\alpha[\e^{ip^{\mu}x_{\mu}} \xi_\alpha(\mathbf{p}) a_\alpha(\mathbf{p})+ \e^{-ip^{\mu}x_{\mu}} \zeta_\alpha(\mathbf{p}) b_\alpha^{\dagger}(\mathbf{p})],
\end{equation}

\noindent where the index $\alpha$ takes the values $\{+,-\},\{-,+\}$.  The spinors $\xi_\alpha(\mathbf{p})$ and $\zeta_\alpha(\mathbf{p})$ are of the general form

\begin{equation}
\label{eqn:genelko}
\chi(\mathbf{p})=\left(\begin{array}{cccc}
\eta\Theta\phi^*(\mathbf{p})\\
\phi(\mathbf{p})
\end{array}\right),
\end{equation}

\noindent where $\eta$ is a nonzero complex number\footnote{We must have $|\eta|= 1$ in order for $\xi^\dagger\xi$ to have its usual interpretation as a probability density.},  and

\begin{eqnarray}
 \phi(\mathbf{p})= \kappa^-\phi(\mathbf{0}).
\end{eqnarray}

\noindent It follows easily that

\begin{eqnarray}
 \chi(\mathbf{p})= \kappa \chi(\mathbf{0}).
\end{eqnarray}

\noindent The spinor $\phi(\mathbf{0})$ is chosen\footnote{We do not need explicit formulas for the Elko rest spinors.  They may be found in \cite[Sec.~2.2]{ahluwalia2010elko}.} so that the $\phi(\mathbf{p})$ satisfy the equations

\begin{equation}\label{eqn:right}
 \boldsymbol{\sigma}\cdot\hat{\mathbf{p}}[\phi(\mathbf{p})]=\pm[\phi(\mathbf{p})],
\end{equation}
\begin{equation}\label{eqn:left}
 \boldsymbol{\sigma}\cdot\hat{\mathbf{p}}[\eta\Theta\phi^*(\mathbf{p})]=\mp[\eta\Theta\phi^*(\mathbf{p})].
\end{equation}

\noindent This means that the two-component vectors formed from the top two components and the bottom two components of $\chi(\mathbf{p})$ have opposite helicities.  For this reason, $\alpha$ is termed a dual-helicity index in \cite[Sec.~3]{AG05}.
The Elko spinors $\xi_\alpha(\mathbf{p})$ and $\zeta_\alpha(\mathbf{p})$ are eigenspinors with eigenvalues $\pm 1$ not of the parity operator $P$, but of the charge conjugation operator $C$.

A new dual $\neg$ is defined for the Elko field (see \cite[Sec.~2.3]{ahluwalia2010elko}).  The propagator turns out to be the Klein-Gordon propagator in the absence of a preferred direction, but in general there is an extra term \cite[App.~A.2]{ALS09b}.   The mass dimensionality of the Elko field is one.
This severely restricts the possible interactions of the Elko field with Standard Model matter.  Hence the Elko field is a plausible candidate for a dark matter field.

\section{Weinberg's definition of a quantum field}
\label{sec:weinberg}

Weinberg gives a careful definition of a quantum field which goes beyond just writing down formulae for spinors.  Ahluwalia describes some of the benefits of this approach in \cite[pp2--3]{Ahl09}.  To formulate Weinberg's theory, we need to describe the mathematical setting in which he works; see \cite[Sec.~5.1]{Wei95} for more details.  Consider a physical system with Hilbert state space $H$ and a unitary representation $U(\Lambda,a)$ of ${\mathcal P}^0$.  By an {\em operator} we mean a linear operator from $H$ to $H$; we denote the space of operators by $L(H)$.  Below we consider various operators and operator-valued functions such as the interaction $V(t)$, the Hamiltonian density $\mathcal{H}(x)$, the creation and annihilation operators $a^\dagger(\mathbf{p},\sigma)$ and $a(\mathbf{p},\sigma)$, and the $U(\Lambda,a)$ themselves.  A Weinberg quantum field is an array of operator-valued functions $\Psi(x)= \Psi_i(x)$ labelled by an index $i$ and having certain transformation properties (see below).

Weinberg argues that quantum fields take the form they do because the $S$-matrix must be Poincar\'e-invariant and satisfy the Cluster Decomposition Principle.
Lorentz invariance of the $S$-matrix is guaranteed if the interaction can be written as
\begin{eqnarray}
V(t)=\int d^3\mathbf{x}\mathcal{H}(\mathbf{x},t),
\end{eqnarray}
where the Hamiltonian density $\mathcal{H}(x)$ has two properties: first, it obeys  the scalar transformation law
\begin{eqnarray}
U(\Lambda,a)\mathcal{H}(x)U(\Lambda,a)^{-1}=\mathcal{H}(\Lambda x+a),
\end{eqnarray}
and second, it commutes with itself at spacelike separation---that is,
\begin{eqnarray}
 \mathcal{H}(x)\mathcal{H}(y)= \mathcal{H}(y)\mathcal{H}(x)
\end{eqnarray}
if $x-y$ is spacelike.  The second property is called locality.  The point of locality is to prevent problems with time-ordering.

We now spell out Weinberg's definition of a quantum field.  The ingredients we need are the following.  We consider massive particles with positive energy and mass $m$.  We take $H$ and $U(\Lambda,a)$ to be as given in Section~\ref{sec:reps}, for some choice of irreducible representation $R(\Lambda)$ of ${\rm SO}(3)$ of spin $s$.  Let $D(\Lambda)$ be a $t$-dimensional representation of $\mathcal{L}^0$ for some positive integer $t$.  We define a \textit{Weinberg quantum field based on the data} $(H,R(\Lambda),U(\Lambda,a),D(\Lambda))$\footnote{The data $R(\Lambda)$ and $U(\Lambda,a)$ are not independent---each determines the other---but we include them both for emphasis.} to be a collection of functions $\Psi(x)=(\Psi_i(x))_{1\leq i\leq t}$ from $\mathbb{R}^4$ to $L(H)$ such that for all $(\Lambda,a)\in\mathcal{P}^0$, we have
\begin{equation}
\label{eqn:fieldeqn}
U(\Lambda,a)\Psi_i(x)U(\Lambda,a)^{-1}=\sum_jD_{ij}(\Lambda^{-1})\Psi_j(\Lambda x+a).
\end{equation}
We say that a Weinberg quantum field---or, more generally, a collection of Weinberg quantum fields---is \textit{local} if for any $\Psi$ and $\Phi$ in the collection, for any indices $i$ and $j$ and for any $x,y\in\mathbb{R}^4$ such that $x-y$ is spacelike, the field components $\Psi_i(x)$ and $\Phi_j(y)$ commute (or anti-commute if the fields are both fermionic).  The point of this definition is that we can build up a Hamiltonian density ${\mathcal H}(x)$ satisfying the desired properties from local Weinberg quantum fields $\Psi^1(x),\ldots, \Psi^N(x)$: we set
\begin{eqnarray}
\mathcal{H}(x)=\sum_N\sum_{\ell_1\cdot\cdot\cdot\ell_N}g_{\ell_1\cdot\cdot\cdot\ell_N}\Psi^1_{\ell_1}(x)\cdot\cdot\cdot\Psi^N_{\ell_N}(x)
\end{eqnarray}
for suitably transforming quantities $g_{\ell_1\cdot\cdot\cdot\ell_N}$.

If one assumes that the Cluster Decomposition Principle holds then it follows from quite general arguments \cite[p197]{Wei95} that $\Psi(x)$ can be written in the form
\begin{equation}
\label{eqn:weifld}
\Psi(x)=\int d^3{\mathbf{p}}\sum_{\sigma}\left[u(x;\mathbf{p},\sigma)a(\mathbf{p},\sigma)+ v(x;\mathbf{p},\sigma)b^{\dagger}(\mathbf{p},\sigma)\right],
\end{equation}
where each component $u_i(x;\mathbf{p},\sigma)$ and $v_i(x;\mathbf{p},\sigma)$ is a complex-valued function of $x$, $\mathbf{p}$ and $\sigma$.  To find the possible Weinberg quantum fields based on given data $(H,R(\Lambda),U(\Lambda,a),D(\Lambda))$ explicitly, we need to determine the possible coefficient functions $u(x;\mathbf{p},\sigma)$ and $v(x;\mathbf{p},\sigma)$.

One can deduce the functional dependence of the $u(x;\mathbf{p},\sigma)$ and $v(x;\mathbf{p},\sigma)$ from eq.~(\ref{eqn:fieldeqn}): one multiplies eq.~(\ref{eqn:fieldeqn}) on the left by $U(\Lambda,a)$ and on the right by $U(\Lambda,a)^{-1}$, then uses \cite[eq.~(4.2.12)]{Wei95} to evaluate $U(\Lambda,a)a^{\dagger}(\mathbf{p},\sigma)U(\Lambda,a)^{-1}$ and $U(\Lambda,a)a(\mathbf{p},\sigma)U(\Lambda,a)^{-1}$
in the RHS of the resulting equation.\footnote{Note that for fixed $x$, ${\mathbf p}$ and $\sigma$, all of the quantities in the integrand on the RHS of eq.~(\ref{eqn:weifld}) apart from the creation and annihilation operators are c-numbers, so they commute with $U(\Lambda,a)$.}  By taking $(\Lambda,a)$ to be a spacetime displacement ($\Lambda= I$), one deduces that $u(x;\mathbf{p},\sigma)$ and $v(x;\mathbf{p},\sigma)$ are of the form
\begin{equation}
 u(x;\mathbf{p},\sigma)= \frac{1}{(2\pi)^{\frac{3}{2}}}\e^{ip^{\mu}x_{\mu}}u(\mathbf{p},\sigma)
\end{equation}
and
\begin{equation}
 v(x;\mathbf{p},\sigma)= \frac{1}{(2\pi)^{\frac{3}{2}}}\e^{-ip^{\mu}x_{\mu}}v(\mathbf{p},\sigma),
\end{equation}
so we can write $\Psi(x)$ as
\begin{equation}
\label{eqn:weinox}
\Psi(x)=\int \frac{d^3{\mathbf{p}}}{(2\pi)^{\frac{3}{2}}} \sum_{\sigma}\left[\e^{ip^{\mu}x_{\mu}}u(\mathbf{p},\sigma)a(\mathbf{p},\sigma)+ \e^{-ip^{\mu}x_{\mu}}v(\mathbf{p},\sigma)b^{\dagger}(\mathbf{p},\sigma)\right].
\end{equation}
The $u(\mathbf{p},\sigma)$ and $v(\mathbf{p},\sigma)$ are the spinors in the usual theory, although Weinberg avoids this terminology in \cite{Wei95}.  By considering Lorentz boosts, one can show that $u(\mathbf{p},\sigma)$ is completely determined for all $\mathbf{p}$ by its values for the rest spinors $u(\mathbf{0},\sigma)$, and likewise for $v(\mathbf{p},\sigma)$.

\section{The Dirac and Elko fields in Weinberg's formalism}
\label{sec:diracelko}

In this section we investigate whether the Elko field can be interpreted as a quantum field in the sense of Weinberg if we take the symmetry group to be the full Poincar\'e group.  Initially we assume, as in \cite{AG05}, that the finite-dimensional representation $D(\Lambda)$ is the chiral representation $D^{\rm ch}(\Lambda)$ from Section~\ref{sec:elkos} \nfs  The state space $H_1$ and the unitary representation $U(\Lambda,a)$ of ${\mathcal P}^0$ are not given explicitly in \cite{AG05}, so we need to specify them.  Under the assumption that $U(\Lambda,a)$ is irreducible, $H_1$ and $U(\Lambda,a)$ must be of the form given in Section~\ref{sec:reps} for some spin $s$ and some irreducible representation $R(\Lambda)$ of ${\rm SO}(3)$.  The derivation below shows that $s$ must be half: this need not be assumed {\em a priori}.

We also need to relate the quantities on the RHS of eq.~(\ref{eqn:localelko}) to those on the RHS of eq.~(\ref{eqn:weinox}).  According to the recipe in Section~\ref{sec:weinberg}, we must identify the Elko field index $\alpha$ with the state space index $\sigma$, which labels the basis vectors of some representation $R(\Lambda)$ of the little group ${\rm SO}(3)$.  We identify $a_\alpha(\mathbf{p})$ with the annihilation operator $a(\mathbf{p},\sigma)$, and likewise we identify $b^\dagger_\alpha(\mathbf{p})$ with $b^\dagger(\mathbf{p},\sigma)$.
Then we have reduced our problem to the following question: for given representations $R(\Lambda)$ and $U(\Lambda,a)$, is the Elko field eq.~(\ref{eqn:localelko})
a solution to eq.~(\ref{eqn:fieldeqn})?  It turns out that it is enough to consider the rest spinors only.

To answer this question, we first consider a special case.  Define $R^{\rm std}(\Lambda)$ to be the spin-half representation of ${\rm SO}(3)$ such that the corresponding generators of angular momentum are given by $\boldsymbol{J}= \frac{1}{2}\boldsymbol{\sigma}$, where the $\sigma_i$ are the Pauli matrices.  Weinberg shows that the Dirac field eq.~(\ref{eqn:diracfld}) is essentially the only Weinberg quantum field based on the data $(H,R^{\rm std}(\Lambda),U(\Lambda,a),D^{\rm ch}(\Lambda))$.  Since the Elko field and the Dirac field are not the same, it follows that the Elko field is not a Weinberg quantum field based on the data $(H,R^{\rm std}(\Lambda),U(\Lambda,a),D^{\rm ch}(\Lambda))$.  Below we recall the relevant parts of his derivation (see \cite[Sec.~5.5]{Wei95}).

Let $\boldsymbol{J}$ be the generators of angular momentum corresponding to the representation $R(\Lambda)$ of ${\rm SO}(3)$.  Each of the three components of $\boldsymbol{J}$ is a $(2s+1)\times (2s+1)$ matrix.  Relabel the components $u_i(\mathbf{0},\sigma)$ of the rest spinors as $u_{m\pm}(\mathbf{0},\sigma)$, where $m$ takes the values $\pm$ and $++$, $-+$, $+-$, $--$ correspond to $i=1,2,3,4$ respectively.  Relabel the components $v_i(\mathbf{0},\sigma)$ as $v_{m\pm}(\mathbf{0},\sigma)$ similarly.  Now define $2\times (2s+1)$ matrices $U_\pm$, $V_\pm$ by
\begin{eqnarray}
\label{eqn:UuVv}
 (U_\pm)_{m\sigma}= u_{m\pm}(\mathbf{0},\sigma),\ (V_\pm)_{m\sigma}= v_{m\pm}(\mathbf{0},\sigma).
\end{eqnarray}

\noindent It follows from eq.~(\ref{eqn:fieldeqn}) that the matrices $U_\pm$, $V_\pm$ satisfy the equations
\begin{eqnarray}
\label{eqn:spinormatrixequation}
U_+\boldsymbol{J}=\frac{1}{2}\boldsymbol{\sigma}U_+,\quad
U_-\boldsymbol{J}=\frac{1}{2}\boldsymbol{\sigma}U_-,\quad
V_+\boldsymbol{J}^*=-\frac{1}{2}\boldsymbol{\sigma}V_+,\quad
V_-\boldsymbol{J}^*=-\frac{1}{2}\boldsymbol{\sigma}V_-;
\end{eqnarray}
see \cite[eqs.~(5.5.3) and (5.5.4)]{Wei95}.
By Schur's lemma, we must have $s=\frac{1}{2}$ and $\boldsymbol{J}$ must be the same as $\frac{1}{2}\boldsymbol{\sigma}$ up to a similarity transformation.
Explicitly, eq.~(\ref{eqn:UuVv}) gives

\begin{eqnarray}
\label{eqn:U+}
U_+=\left(\begin{array}{cccc}
u_1(\mathbf{0},{\textstyle \frac{1}{2}})&u_1(\mathbf{0},-{\textstyle \frac{1}{2}})\\
u_2(\mathbf{0},{\textstyle \frac{1}{2}})&u_2(\mathbf{0},-{\textstyle \frac{1}{2}})
\end{array}\right)
\end{eqnarray}
\begin{eqnarray}
\label{eqn:U-}
U_-=\left(\begin{array}{cccc}
u_3(\mathbf{0},{\textstyle \frac{1}{2}})&u_3(\mathbf{0},-{\textstyle \frac{1}{2}})\\
u_4(\mathbf{0},{\textstyle \frac{1}{2}})&u_4(\mathbf{0},-{\textstyle \frac{1}{2}})
\end{array}\right)
\end{eqnarray}
\begin{eqnarray}
\label{eqn:V+}
V_+=\left(\begin{array}{cccc}
v_1(\mathbf{0},{\textstyle \frac{1}{2}})&v_1(\mathbf{0},-{\textstyle \frac{1}{2}})\\
v_2(\mathbf{0},{\textstyle \frac{1}{2}})&v_2(\mathbf{0},-{\textstyle \frac{1}{2}})
\end{array}\right)
\end{eqnarray}
\begin{eqnarray}
\label{eqn:V-}
V_-=\left(\begin{array}{cccc}
v_3(\mathbf{0},{\textstyle \frac{1}{2}})&v_3(\mathbf{0},-{\textstyle \frac{1}{2}})\\
v_4(\mathbf{0},{\textstyle \frac{1}{2}})&v_4(\mathbf{0},-{\textstyle \frac{1}{2}})
\end{array}\right),
\end{eqnarray}

\noindent where $m$ positive (resp.\ negative) labels row 1 (resp.\ 2), and $\sigma$ positive (resp.\ negative) labels column 1 (resp.\ 2).

Suppose we choose $\boldsymbol{J}$ to be equal to $\frac{1}{2}\boldsymbol{\sigma}$: that is, suppose we choose $R(\Lambda)$ to be $R^{\rm std}(\Lambda)$.  It then follows from Schur's lemma that the $U_{\pm}$ matrices must be proportional to the identity and the $V_{\pm}$ matrices must be proportional to $\sigma_2$, so we have:
\begin{eqnarray}
\label{eqn:cdpm}
U_+=\left(\begin{array}{cccc}
c_+&0\\
0&c_+
\end{array}\right),\quad
U_-=\left(\begin{array}{cccc}
c_-&0\\
0&c_-
\end{array}\right),\quad
V_+=\left(\begin{array}{cccc}
0&-d_+\\
d_+&0
\end{array}\right),\quad
V_-=\left(\begin{array}{cccc}
0&-d_-\\
d_-&0
\end{array}\right)
\end{eqnarray}
for some constants $c_\pm$ and $d_\pm$.  Equations~(\ref{eqn:U+})--(\ref{eqn:cdpm}) imply that the rest spinors are given by
\begin{eqnarray}
\label{eqn:gendiracrest}
u(\mathbf{0},{\textstyle \frac{1}{2}})=\left(\begin{array}{cccc}
c_+\\
0\\
c_-\\
0
\end{array}\right),\quad u(\mathbf{0},-{\textstyle \frac{1}{2}})=\left(\begin{array}{cccc}
0\\
c_+\\
0\\
c_-
\end{array}\right),\quad
v(\mathbf{0},{\textstyle \frac{1}{2}})=\left(\begin{array}{cccc}
0\\
d_+\\
0\\
d_-
\end{array}\right),\quad v(\mathbf{0},-{\textstyle \frac{1}{2}})=\left(\begin{array}{cccc}
-d_+\\
0\\
-d_-\\
0
\end{array}\right).
\end{eqnarray}

A further analysis involving locality and the extended Poincar\'e group (see below) allows one to determine the value of the constants $c_\pm$, $d_\pm$.  One finds that the resulting rest spinors are precisely the Dirac rest spinors from eqs.~(\ref{eqn:diracrestA}) and (\ref{eqn:diracrestB}).  Hence the Dirac field is the only local Weinberg quantum field based on the data $(H,R^{\rm std}(\Lambda),U(\Lambda,a),D^{\rm ch}(\Lambda))$.

Now we return to the more general case, in which we replace the representation $R^{\rm std}(\Lambda)$ with another representation $R(\Lambda)$.\footnote{We can identify the index $\alpha$ not with the usual basis of $R^{\rm std}(\Lambda)$---the one labelled by $\sigma$---but some other arbitrary basis.  This amounts to replacing $R^{\rm std}(\Lambda)$ with another representation $R(\Lambda)$, so this case is covered by the present argument.}  For good measure, let us also allow the representation $D(\Lambda)$ to be not the chiral representation $D^{\rm ch}(\Lambda)$, but another representation in the same isomorphism class. The angular momentum $\boldsymbol{M}$ corresponding to $D(\Lambda)$ is related to $\frac{1}{2}\boldsymbol{\sigma}$ by a similarity transform.  Equation~(\ref{eqn:spinormatrixequation}) becomes
\begin{eqnarray}\label{eqn:UVeqns}
U_+\boldsymbol{J}=\boldsymbol{M}U_+,\quad U_-\boldsymbol{J}=\boldsymbol{M}U_-,\quad
V_+\boldsymbol{J}^*=-\boldsymbol{M}V_+,\quad V_-\boldsymbol{J}^*=-\boldsymbol{M}V_-,
\end{eqnarray}
where $\boldsymbol{J}$ is the angular momentum corresponding to $R(\Lambda)$.  It follows again from Schur's Lemma that $s= \frac{1}{2}$, that $\boldsymbol{J}$ is related to $\boldsymbol{M}$ (and hence to $\boldsymbol{\sigma}$) by a similarity transform, that $U_+$ and $U_-$ are proportional and that $V_+$ and $V_-$ are proportional: say,
\begin{eqnarray}
\label{eqn:UVschur}
 U_+=AU_-,\quad V_+=BV_-
\end{eqnarray}

\noindent for some scalars $A$ and $B$.

Suppose we have a solution to eq.~(\ref{eqn:UVeqns}) such that each of $u(\mathbf{0},{\textstyle \frac{1}{2}})$, $u(\mathbf{0},{\textstyle -\frac{1}{2}})$, $v(\mathbf{0},{\textstyle \frac{1}{2}})$, $v(\mathbf{0},{\textstyle -\frac{1}{2}})$ is of the form in eq.~(\ref{eqn:genelko}) for some $\phi$.  Then
\begin{eqnarray}
\label{eqn:arbrest}
u(\mathbf{0},{\textstyle \frac{1}{2}})=\left(\begin{array}{cccc}
-\eta b_1^*\\
\eta a_1^*\\
a_1\\
b_1
\end{array}\right),
u(\mathbf{0},{\textstyle -\frac{1}{2}})=\left(\begin{array}{cccc}
-\eta b_2^*\\
\eta a_2^*\\
a_2\\
b_2
\end{array}\right),
v(\mathbf{0},{\textstyle \frac{1}{2}})=\left(\begin{array}{cccc}
-\eta d_1^*\\
\eta c_1^*\\
c_1\\
d_1
\end{array}\right),
u(\mathbf{0},{\textstyle -\frac{1}{2}})=\left(\begin{array}{cccc}
-\eta d_2^*\\
\eta c_2^*\\
c_2\\
d_2
\end{array}\right)
\end{eqnarray}
for some constants $a_1$, $b_1$, $a_2$, $b_2$, $c_1$, $d_1$, $c_2$, $d_2$ (recall from Section~\ref{sec:elkos} that the Elko spinors are a special case of this).  Equations~(\ref{eqn:UVschur})--(\ref{eqn:arbrest}) and (\ref{eqn:U+})--(\ref{eqn:V-}) imply that
\begin{eqnarray}\label{eqn:abcd}
\left(\begin{array}{cccc}
-\eta b_1^*&-\eta b_2^*\\
\eta a_1^*&\eta a_2^*
\end{array}\right)=A\left(\begin{array}{cccc}
a_1&a_2\\
b_1&b_2
\end{array}\right),\quad \left(\begin{array}{cccc}
-\eta d_1^*&-\eta d_2^*\\
\eta c_1^*&\eta c_2^*
\end{array}\right)=B\left(\begin{array}{cccc}
c_1&c_2\\
d_1&d_2
\end{array}\right).
\end{eqnarray}
Each column in the two matrix equations in eq.~(\ref{eqn:abcd}) yields a pair of equations; taking their ratios gives the equations
\begin{equation}
a_1a_1^*=-b_1b_1^*,\quad a_2a_2^*=-b_2b_2^*,\quad c_1c_1^*=-d_1d_1^*,\quad c_2c_2^*=-d_2d_2^*,
\end{equation}
which have only the trivial solution. In particular, the Elko rest spinors $\xi_\alpha(\mathbf{0})$ and $\zeta_\alpha(\mathbf{0})$ cannot be the rest spinors of a Weinberg quantum field.  It follows that the Elko field is not a Weinberg quantum field based on the data $(H,R(\Lambda),U(\Lambda,a),D(\Lambda))$, at least not when interpreted as above.  This is the main conclusion of the present work.  The same arguments hold for the Elko fields with spinors $\lambda(\mathbf{p})$ and $\rho(\mathbf{p})$ given in \cite[eq.~(3.3)]{AG05}.

Note that the assumption of locality was not needed in the above argument.  Locality allows one to pin down the form of the spinors $u(\mathbf{p},\sigma)$ and $v(\mathbf{p},\sigma)$ in the Dirac field by determining the values of $c_\pm$ and $d_\pm$ in eq.~(\ref{eqn:gendiracrest}).  We briefly recall the rest of this argument, which involves the transformation properties under the extended Lorentz and Poincar\'e groups.  The representation $U(\Lambda,a)$ of ${\mathcal P}^0$ on $H_1$ can be extended to give a representation of ${\mathcal P}$: for instance, the operator $U({\sf P})$ is multiplication by a phase.  This gives a representation of ${\mathcal P}$ on the total state space $H$.  Weinberg assumes that the overall Hamiltonian density ${\mathcal H}(x)$ is parity-invariant.  Since the Dirac field $\psi(x)$ appears in the Hamiltonian density, its parity transform $U(P)\psi(x)U(P)^{-1}$ also does.  Locality requires that $\psi(x)$ and $U(P)\psi(x)U(P)^{-1}$ commute (see \cite[pp221--222]{Wei95}), and this---together with the requirement that $\psi(x)$ is local---determines the values of $c_\pm$ and $d_\pm$ (up to an overall phase and the freedom to replace $\psi(x)$ with $\gamma_5\psi(x)$).

\section{Non-standard Wigner classes}
\label{sec:nonstd}

Wigner described the possible irreducible unitary representations of the strict Poincar\'e group in 1939 \cite{Wig39}: they are the representations $U(\Lambda,a)$ of ${\mathcal P}^0$ on $H_1$ given in Section~\ref{sec:reps} \nfs  Later he extended this work to give a classification of the irreducible unitary representations $U(\Lambda,a)$ of the extended Poincar\'e group ${\mathcal P}$ \cite{Wig64}, \cite[Sec.~2.C]{Wei95}.  There are four isomorphism classes of representations: one so-called standard Wigner class and three non-standard Wigner classes.  The standard Wigner class is the representation of ${\mathcal P}$ on $H_1$ discussed at the end of Section~\ref{sec:diracelko} \nfs  The state space $H_1$ in the non-standard cases is different from the one described in Section~\ref{sec:reps}, as we explain below.

Ahluwalia and Grumiller study the commutation relations of the discrete transformations $C$ and $P$ \cite[eq.~(4.16)]{AG05}.  Their results show that the finite-dimensional representation $D^{\rm ch}(\Lambda)$ has a structure closely analogous to that of one of the non-standard Wigner classes (see \cite[p4]{AG05} for discussion).  Motivated by this and by results of Ahluwalia, Johnson and Goldman \cite{AJG93}, it is logical to study Weinberg quantum fields based on non-standard Wigner classes and to search for fields of Elko type.  At first glance, this may seem strange: our argument that the Elko field is not a Weinberg quantum field involves eq.~(\ref{eqn:fieldeqn}) applied only to elements of ${\mathcal P}^0$ and ${\mathcal L}^0$, and the discrete symmetries do not appear to play any part.
The explanation for this apparent paradox is as follows.  We assume the one-particle state space $H_1$ carries an irreducible unitary representation of ${\mathcal P}$.  The restriction of the representation to ${\mathcal P}^0$ is isomorphic to a direct sum of irreducible representations of ${\mathcal P}^0$.  If the representation of ${\mathcal P}$ we started with is in the standard Wigner class then this restriction is irreducible: there is only one irreducible summand, namely $H_1$ endowed with the representation $U(\Lambda,a)$ of ${\mathcal P}^0$ from Section~\ref{sec:reps} \nfs  This is the case considered above.  If the representation of ${\mathcal P}$ is in one of the non-standard Wigner classes then the restriction to ${\mathcal P}^0$ is the sum of {\bf two} irreducible representations.  (These turn out to be isomorphic to each other; one may choose basis kets of the form $\left|p,\sigma,\tau\right\rangle$, where $p$ and $\sigma$ are as before and $\tau$ is a degeneracy index which distinguishes between the irreducible components.\footnote{Time reversal $U({\sf T})$ couples states with different values of $\tau$.})  Hence $H_1$ has a different mathematical structure in the non-standard cases, even when we consider only representations of ${\mathcal P}^0$.  We believe these non- standard cases, which are not worked out in detail in \cite{Wei95}, are worth further study; even if one cannot find Elko-type fields in this setting, perhaps there are other as yet unexplored Weinberg quantum fields that are candidates for dark matter.  The authors will investigate this in forthcoming work.

We finish with some remarks on work of Lee and Wick which is relevant here.  According to \cite{LW71}, if a field is local then the underlying representation of ${\mathcal P}$ must come from the standard Wigner class. In their work, however, one allows oneself the freedom to multiply the original $U({\sf P})$ and $U({\sf T})$ by symmetries of the internal state space.  For the non-standard Wigner classes, one would expect there to be plenty of internal symmetries because of the extra degrees of freedom coming from the index $\tau$.  A full study of the possible Weinberg quantum fields would involve an investigation of these internal symmetries.

\section{Conclusion}
\label{sec:conclusion}

We have shown that the Elko field does not transform covariantly under the full Poincar\'e group, supporting the results of Ahluwalia, Lee and Schritt \cite{ahluwalia2010elko}\cite{ALS09b}.  Ahluwalia \cite{Ahl09} and Ahluwalia and Horvath \cite{AH} argue that the postulate of Poincar\'e covariance is based on experimental evidence involving standard model matter. We do not know whether rods and clocks made of dark matter would respect the same symmetries, hence the symmetry group of a dark matter field need not be the Poincar\'e group \cite{Ahl09}.  A natural next step would be to construct an analogue of Weinberg's formalism with the Poincar\'e group replaced by another symmetry group such as SIM(2).

\section{Acknowledgements}

We are grateful to the Dark2009 conference organisers for giving us the opportunity to present an earlier version of this work \cite{Conferenceproceedings}.  Most of the ideas in this paper had their roots in discussions of the authors with Dharamvir Ahluwalia, Cheng-Yang Lee, Dimitri Schritt and Thomas Watson, and we thank them for their contributions and their encouragement.  In particular, we thank Ahluwalia for reading an earlier draft of this work and pointing out some mistakes.
The second author thanks Ahluwalia for introducing him to the Elko field and to quantum field theory in general.

We thank the referees for their careful reading of the paper and for supplying a number of corrections and suggestions for improvement.

\bibliography{elkoweinberg3}

\end{document}